\documentclass[a4paper,11pt]{article}
\usepackage{a4wide}
\usepackage{amsmath,amssymb}

\usepackage{color}

\numberwithin{equation}{section}
\usepackage{graphicx}
\usepackage{cite}

\let\mathscr\mathcal
\def\C{\mathbb{C}}

\def\R{\mathbb{R}}

\def\CP{\mathbb{CP}}
\def\N{\mathscr{N}}

\def\cF{\mathcal{F}}
\def\cA{\mathscr{A}}
\def\cB{\mathcal{B}}

\def\Z{\mathbb{Z}}

\def\dP{dP}

\def\div{\mathrm{div}}

\DeclareMathOperator{\tr}{tr}

\DeclareMathOperator{\diag}{diag}

\def\five-dimensional{5D}
\def\four-dimensional{4D}

\def\AdS{{\text{AdS}}}
\def\vol{\mathrm{vol}}
\def\CFT{\text{CFT}}

\def\contract#1#2{(#1\cdot#2)}
\def\vev#1{\langle#1\rangle}

\def\f{\mathcal{B}}

\begin{document}
\begin{titlepage}
\begin{flushright}
MCTP-05-104\\
 MIT-CTP-3740\\
NSF-KITP-05-117\\
UT-05-21\\
hep-th/0601054
\end{flushright}

\vskip 1cm

\begin{center}
\textbf{\LARGE
Triangle Anomalies from Einstein Manifolds}\\

\vskip1cm

{\Large Sergio Benvenuti$^1$, Leopoldo A. Pando Zayas$^{2,4}$ \\[2mm]
 and Yuji Tachikawa$^{3,4}$}\\

\vskip1cm

\textit {$^1$\ { Scuola Normale Superiore, Pisa,\\
    and INFN, Sezione di Pisa, Italy.}}\\

\vskip.5cm

\textit{$^2$\ { Michigan Center for Theoretical Physics,
 Randall Laboratory of Physics, \\
 The University of Michigan Ann Arbor, MI48109-1040,USA }}\\

 \vskip.5cm

\textit {$^3$\ { Department of Physics, Faculty of Science,} \\
{University of Tokyo, Tokyo 113-0033,  Japan}}\\

\vskip.5cm

\textit {$^4$\ {Kavli Institute for Theoretical Physics,}}\\
\textit {University of California, Santa Barbara, CA 93106, USA}\\

\vskip 1cm

\vskip.5cm

{\large\textbf{Abstract}}

\end{center}
The triangle anomalies in conformal field theory, which can be
used to determine the central charge $a$, correspond to the
Chern-Simons couplings of gauge fields in $\AdS_5$ under the
gauge/gravity correspondence. We present a simple geometrical
formula for the Chern-Simons couplings in the case of type IIB
supergravity compactified on a five-dimensional Einstein manifold
$X$. When $X$ is a circle bundle over del Pezzo surfaces or a
toric Sasaki-Einstein manifold, we show that the gravity result is
in perfect agreement with the corresponding quiver gauge theory.
Our analysis reveals an interesting connection with the
condensation of giant gravitons or dibaryon operators which
effectively induces a rolling among Sasaki-Einstein vacua.

\vbox{}\vspace{1\fill}

\end{titlepage}



\section{Introduction}
Recent years have seen a tremendous progress in
developing the Anti de Sitter/Conformal Field Theory
($\AdS_5/\CFT_4$) correspondence \cite{Maldacena}. The
correspondence arises from considering a large number $N$ of
D3-branes placed at a singularity, which is locally the tip of a
real cone over a five-dimensional Einstein manifold $X$. It
predicts the equivalence of the field theory on the stack of the
D3-branes and the Type IIB theory on $\AdS_5\times X$.

The very first check of the correspondence involves the symmetries
on the two sides. The conformal group of $\CFT_4$ is
mapped to the isometry group of $\AdS_5$.
Other global symmetries in the $\CFT_4$ are mapped to
gauge symmetries in $\AdS_5$. More precisely, global symmetry
currents $J_I$ on the boundary correspond to massless gauge
fields $A^I$ in the five-dimensional ($5$d) bulk with the boundary coupling
$ \int d^4x \, A^I  J_I.$

The global symmetries in the CFT side in general have triangle anomalies
among them. They are mapped to the Chern-Simons (CS) couplings
$(24\pi^2)^{-1}\int c_{IJK} A^I\wedge F^J\wedge F^K$ for the
$5$d gauge fields, and
the matching between them
provides a quantitative check of the $\AdS/\CFT$ correspondence.
It was carried out in \cite{W} for $X =
S^5$ using supergravity results of \cite{sugra1,sugra2},
but it has not yet been done for other Einstein manifolds.
It is well-known that triangle anomalies can be extracted by a
simple one-loop computation in the gauge theories, and that they are
topological objects.
We thus expect that  it should be possible
to develop a generic quantitative understanding also on the
gravity side of the duality,
because they should belong to ``protected sectors'' of
the $\AdS_5/\CFT_4$ correspondence.

Other types of ``protected sectors'' of the $\AdS/\CFT$ correspondence
are given by the Bogomol'ny-Prasad-Sommerfield (BPS) operators,
which are protected by supersymmetry. In this case one can map the
scaling dimensions of the BPS operators to the energy of the
corresponding BPS states in type IIB string theory on $\AdS_5
\times X$. We can expect it to be possible to understand the
dual BPS objects on the gravity side in general, without the need
of having the explicit metrics. This is indeed the case, for instance,
for dimensions of baryonic BPS operators, corresponding to the volumes
of supersymmetric (SUSY) cycles, which can be computed with the
procedure uncovered in \cite{MSY}. In the same way,
we expect that the CS coefficients can be calculated in the
gravity side without the knowledge of the explicit metrics.

The $5$d Chern-Simons coefficients also appear prominently
in the analysis of M-theory on Calabi-Yau threefolds.
They  are given in terms of the triple
intersections of three four-cycles of the Calabi-Yau. Hence,
we expect to find a similarly robust formula for the CS coefficients
in the case of Type IIB supergravity on compact, positively
curved, Einstein manifolds $X$.

Thus, our first objective is to obtain a  geometrical formula for
the Chern-Simons coefficients $c_{IJK}$
for Type IIB supergravity on $\AdS_5 \times X$.
The result we will obtain is so elegant that we would like to give the
formula here.  It is given by 
\begin{equation}\label{finalr}
c_{IJK}=\frac{N^2}{2} \int_X \omega_{\{I}  \wedge \iota_{k_J}
\omega_{K\}}\,.
\end{equation} Here, $N$ is the number of the  flux of the self-dual five-form $F_5$
through $X$, and three-forms $\omega_I$ of $X$ appears in the fluctuation
of $F_5$ via  \begin{equation}
\delta F_5=N(A^I\wedge \omega_I)
\end{equation}where $A^I$ are gauge fields on $\AdS_5$.
Therefore $\omega_I$ determines the
the distribution of $F^I$ in the internal manifold and thus is usually called
wave function.

The Killing vectors $k_I$ measure the non-closedness
of $\omega_I$
by the relation \begin{equation}
d\omega_I+\iota_{k_I}\vol^\circ=0
\end{equation}where $\vol^\circ$ is the volume form normalized to have
$\int_{X}\vol^\circ=1$. The index $I$ runs from $1$ to $d=\ell+b^3$
where $\ell$ is the number of isometries of $X$ and $b^3$ is the
third Betti number of $X^5$.

We will show that this formula
gives robust topological quantities in a
precise sense. In particular,  explicit
knowledge of the Einstein metric on $X$
is not necessary to evaluate the formula (\ref{finalr}).

While our formula (\ref{finalr}) is valid for any Einstein
manifold $X$, the case when $X$ is Sasaki-Einstein (SE) is
especially interesting. In this case, by definition, the cone over
$X$ is Calabi-Yau. Then, minimal $\mathcal{N}=1$ supersymmetry is
preserved and we get an $\mathcal{N}=1$ superconformal field
theory (SCFT).
 Other than the round $S^5$, the only
 Sasaki-Einstein space with an explicitly known metric was $T^{1,1}$
 for a long time; its SCFT dual was first
studied by Klebanov and Witten in \cite{KW}. We now have a
countably infinite number of explicit SE metrics
\cite{Ypq,Lpqr,Martelli:2005wy} and the corresponding quiver gauge
theories \cite{YpqQuiver,geo,LpqQuiver0,LpqQuiver}.
There is a nice
interaction between `topological' objects and objects protected by
supersymmetry.
Thus, we have many examples to test our formula against
field theory expectations.

For generic $4$d $\mathcal{N}=1$ SCFTs, triangle anomalies encode
a lot of physical information and are related to various important
correlators of the symmetry currents and the energy-momentum
tensor \cite{AFGJ}. In particular, the supersymmetric partner of
the energy-momentum tensor is an Abelian global symmetry, which is
called the $R$-symmetry. One important point in $\AdS/\CFT$
correspondence is that the triangle anomaly of the $R$-symmetry,
$c_{RRR}$, which is the central charge $a$ of the SCFT
\cite{AFGJ}, is inversely proportional to the volume of the SE
manifold  \cite{HS,Gubser}.

Quantitative analysis on the field theory side can be done thanks
to `$a$-maximization' \cite{amax}, which determines the
$R$-symmetry. On the gravity side the $R$-symmetry is mapped to
the so-called `Reeb vector' of the internal manifold $X$. In the
case $X$ is  \emph{toric} Sasaki-Einstein, \emph{i.e.}~the
isometry group of $X$ contains a $U(1)^3$, `$Z$-minimization'
\cite{MSY} determines the Reeb vector; it is thus possible to
compare the volume of $X$ and of SUSY $3$-cycles with gauge
theory, as was done in \cite{BZ}. In the case of the recently
found $Y^{p,q}$ and $L^{p,q,r}$, checks of the duality have been
given for $c_{RRR}$ in \cite{YpqQuiver, BBC}, for BPS mesonic
operators in \cite{Benvenuti:2005cz,geo,Kihara:2005nt,Oota:2005mr}
and for various SUSY branes in \cite{CPRV}. There are
also works which clarify the relation between $a$-maximization and
$Z$-minimization through 5d theory in $\AdS_5$
\cite{adsamax,currentcorrelators}. Note that their results are
valid also in the non-toric case.

We enlarge this impressive list of checks of the correspondence
by providing an
explicit evaluation of $c_{IJK}$, through (\ref{finalr}), for
large sets of Sasaki-Einstein manifolds, namely,
circle bundles over del Pezzo surfaces
and toric Sasaki-Einstein manifolds. The evaluation utilizes the flow
triggered by the condensation of the giant gravitons. We analyze
the field theory side using the same flow by the Higgsing using
the dibaryon operators, and we find complete agreement on the
gravity side and the field theory side. For toric SE we obtain
\begin{equation}\label{trca}
c_{IJK} = \frac{N^2}{2}|\det (k_I, k_J, k_K)|
\end{equation} where $k_I \in\Z^3$ is the $I$-th  generator
of the toric cone.  We also call $k_I$ the toric data,
as is customary in string theory literature. In
other words, $c_{IJK}$ is simply given by the area of a triangle
formed by the three toric data.
We recover the formula (\ref{trca}) from field
theory, thus providing a very general check of $\AdS/\CFT$.

We will also analyze the BPS operators which are related to
giant gravitons, emphasizing the interplay between objects
protected by SUSY and topological properties of $X$. Throughout
the analysis, we will see that there is an intricate mixing of the
angular momenta and baryonic charges, which reflects the fact that
the D3-branes wrapping three-cycles in the SE manifold is partly a
giant graviton. This unifies the study of two kind of
supersymmetric states important for the AdS/CFT correspondence.
One is the giant gravitons, corresponding to determinant operators in
$\mathcal{N}=4$ super Yang-Mills,  and
the other is the D3-branes wrapped on supersymmetric
$3$-cycles, corresponding to \emph{dibaryon} operators in the dual
quiver theory.

\vspace{0.9 cm}

 The organization of this paper is the following:
first we sketch in section 2 the supergravity reduction which
gives the formula for the CS terms and gauge coupling constants.
Then, we discuss the normalization of gauge fields and the charges
in section 3, where we will see that the formula for the CS terms
is topological in a precise sense. We evaluate the formulae for
toric Sasaki-Einstein manifolds and for the circle bundles over
del Pezzo surfaces in section 4. In section 5, we turn to field
theory dual and show, based on explicit examples,
 that the results obtained in previous
sections match with predictions based on $\AdS_5/\CFT_4$
correspondence. In section 6, we explain the simplicity of
our results in section 5 using the flow triggered by the
condensation of dibaryons.
 We conclude with some discussions in section 7.
Appendix A contains the detail of the supergravity reduction,
while in appendix B we obtain the triangle anomaly for quiver
theories corresponding to generic toric Sasaki-Einstein manifolds.
Finally in appendix C, we elaborate on the mathematics behind the
charge lattice associated to the five-dimensional Einstein manifold with
isometries.

\section{Perturbative Supergravity Reduction}\label{sugra}

Consider Type IIB theory on $\AdS_5\times X$ where $X$ is an
Einstein manifold of dimension five.
Let us carry out the Kaluza-Klein reduction
and retain only the massless gauge fields. The corresponding
five-dimensional action has  the form
\begin{equation}
S=   \frac12 \int \tau_{IJ}F^I\wedge *F^J
+ \frac1{24\pi^2} \int c_{IJK} A^I\wedge F^J\wedge F^K  +\cdots,\label{5dlagrangian}
\end{equation}which yields
the equation of motion \begin{equation}
\tau_{IJ} d*F^I =\frac{1}{8\pi^2} c_{IJK} F^J \wedge F^K.
\end{equation}
We would like to calculate the Chern-Simons interaction $c_{IJK}$
of the gauge fields. We will eventually choose the indices $I,J,\ldots$
to label the integral basis of the gauge fields in the next
section, but in this section we take them arbitrarily.
We chose the numerical coefficient $(24\pi^2)^{-1}$
so that $c_{IJK}=\tr Q_IQ_JQ_K$ under the AdS/CFT correspondence,
where $Q_I$ is the global symmetry corresponding to the gauge field $A_I$,
and the trace is over the label of Weyl fermions.

The arguments which are to be presented in sections \ref{2.1} and
\ref{2.2} only uses the fact that the metric is Einstein, so it is applicable,
e.g.~to the manifolds $T^{a,b}$ for $(a,b)\ne (1,1)$.

\subsection{The Ansatz and its reduction}\label{2.1}
Since the detail of the reduction is rather tedious,
we present only a rough argument in this section.
Interested readers can consult appendix \ref{details} for the details.
We use three kinds of  Hodge stars, namely
on $X$, on $\AdS_5$ and on $\AdS_5\times X$.
We denote the last one by $*_{10}$,
and the first two by $*$. We hope the context makes clear
which one we used.

The equations of motion  and the Bianchi identity in
Type IIB supergravity are\begin{equation}
R_{\mu\nu}= \frac{c}{24} F_{\mu\alpha\beta\rho\sigma}F_{\nu}{}^{\alpha\beta\rho\sigma},\qquad
F_5=*_{10} F_5,\qquad
dF_5=0
\end{equation}
 where $R_{\mu\nu}$ is the Ricci curvature of the ten-dimensional metric
and $F_5$ is the self-dual five-form field strength. The constant
$c$ depends on conventions.
We set all other form fields and fermions to zero,
and the dilaton to constant
throughout the analysis.

Let $N$ units of five-form flux penetrate $X$, where
we normalize the five-form $F_5$ to have $\int F_5\in2\pi\Z$.
The zero-th order solution is\begin{align}
ds^2 &= L^2ds^2_{\AdS}+L^2ds^2_{X},\\
F_5&= \frac{2\pi N}{V} \left(\vol_X +\vol_{\AdS}\right).
\end{align} where $\vol$ is the volume form  of $X$
and $V=\int_X \vol$.
We take the convention $R_{\mu\nu}=-4g_{\mu\nu}$ for $ds^2_{\AdS}$
and $R_{\mu\nu}=4g_{\mu\nu}$ for $ds^2_{X}$as usual.
$L$ sets the physical length scale.

Suppose $X$ has $\ell$  $U(1)$ isometries $k_a^i$, ($a=1,\ldots,\ell$)
so that $\exp(2\pi k_a^i \partial_i)$ is the identity.   For toric SE manifolds, $\ell=3$.
Let us expand the fluctuation around the zero-th order solution in modes.
One can consistently set to zero  all the modes  which are not
invariant under the $U(1)$ isometries.
We take the usual Kaluza-Klein ansatz for the metric\begin{equation}
ds^2_X=\sum_i (e^i+k_a^iA^a)^2 \label{metricansatz}
\end{equation}where $e^i$ are the f\"unfbein forms of the compact
manifold $X$,
and $A^a$ are  one-forms on $\AdS_5$.

The Ansatz for  $F_5$ is rather intricate already at first order.
We write $F_5$ as the sum of components $F_{p,q}$ which has $p$ legs in $\AdS_5$ and
$q$ legs in $X$ so that \begin{equation}
F_5=F_{0,5}+F_{1,4}+F_{2,3}+F_{3,2}+F_{4,1}+F_{5,0}.
\end{equation}
Then we take the Ansatz to be \begin{align}
F_{0,5}&=\frac{2\pi N}{V} \vol_X, &
F_{5,0}&=\frac{2\pi N}{V} \vol_{\AdS}, \\
F_{1,4}&=\frac{2\pi N}{V} A^a \wedge \iota_{k_a}\vol_X + *F_{4,1}, \label{boo}\\
F_{2,3}&=N F^I \wedge \omega_I, &
F_{3,2}&=N  (*F^I) \wedge *\omega_I.
\label{fluct}
\end{align} Here, $\omega_I$ are three-forms on $X$ to be determined later,
and $F^I$ are two-forms on $\AdS_5$, respectively.
The range in which $I$  can take values is also determined later.
The first term in \eqref{boo} is necessary because eq. \eqref{metricansatz}
modifies the Hodge star.

The exterior derivative is decomposed to $d=d_X + d_\AdS$
where $d_{X,\AdS}$ is the exterior derivative on the respective spaces.
Then, $dF_5=0$ imposes \begin{equation}
d_\AdS F_{p,q+1} + d_X F_{p+1,q}=0 .
\end{equation}
$F_{4,1}$ can be shown to yield massive degrees of freedom, so
we set $F_{4,1}=0$.
Moreover, in order to have massless equation of motion $dF^I=0$  and $d*F^I=0$,
there must be constants $c^a_I$ such that
\begin{equation}
d*\omega_I =0, \qquad
d\omega_I = \frac{2\pi}{V} c^a_I\iota_{k_a}\vol_X, \label{pre}
\end{equation} for $\omega_I$ and\begin{equation}
dA^a=c^a_I F^I\label{mixing}
\end{equation} for $F^I$.
One important property is the non-closedness of $\omega_I$,
which was already pointed out in \cite{currentcorrelators}.
If $d\omega_I=0$ in \eqref{pre},
the allowed number of $F^I$ would be precisely $b^3=\dim H^3(X)$.
The presence of $\iota_{k_a}\vol_X$ enlarges the dimension of
the space of  wavefunctions $\omega_I$ for massless gauge fields
by the number of isometries, $\ell$.
Thus, the index $I$ runs from $1$ to $d$ where \begin{equation}
d=\ell + b^3\label{num}.
\end{equation}
Let us introduce $\vol^\circ\equiv \vol/V$ and
$k_I\equiv 2\pi c^a_I k_a$. Eq. \eqref{pre} becomes
\begin{equation}
d\omega_I + \iota_{k_I}\vol_X^\circ =0. \label{closeduptoisometry}
\end{equation}

We now consider the Chern-Simons couplings. One contribution to the CS
interaction arises as follows.
The Hodge star $*$ for the metric ansatz \eqref{metricansatz}
forces $F_5$ to have a
second-order contribution of the form
\begin{equation}
\delta^{(2)} F\propto A^a \wedge F^I \wedge \iota_{k_a} \omega_I,
\end{equation}just as we had $A^a\wedge \iota_{k_a}\vol_X$ term
in \eqref{boo}.
Then, $d_\AdS F_{3,2} + d_X F_{4,1}=0$
requires the presence of $F^a\wedge F^I$ terms in the right
hand side of the equation of motion.
After combining with the other contribution,
the resulting equation of motion for $F^I$ turns out to be
\begin{equation}
 d*F^I \int_X (\omega_K\wedge *\omega_I+\frac{1}{16V^2}
 \contract{k_K}{k_I}\vol)
=\frac1{8\pi} F^I\wedge F^J \int_X \omega_{\{I} \wedge \iota_{k_J} \omega_{K\}} \label{final}
\end{equation}
where $\contract{a}{b}$ for two one-forms $a=a_idx^i$,
$b=b_i dx^i$ is defined by $\contract{a}{b}=a_i b_j g^{ij}$,
and $\{IJK\}=IJK+IKJ+\cdots$  is the
total symmetrization without $1/6$.
Again, consult appendix \ref{details} for details.

\subsection{Comparison to the 5d Lagrangian}\label{2.2}
Let us write down the formula for $c_{IJK}$ and $\tau_{IJ}$.
In order to determine the combination of $\tau_{IJ}$ and
$c_{IJK}$ entering the five-dimensional action,  we need the normalization
of the kinetic term of $F_5$ entering the ten-dimensional action.
One can resort to string worldsheet perturbation theory,
but there is a quicker way out.
We are normalizing $F_5$ to have $\int F_5\in2\pi\Z$.
Then a D3-brane sources the field $F_5=dC_4$
by the coupling $S= \int_{D3} C_4$.
D3-branes are their own electromagnetic dual, thus
one D3-brane should create five-form flux which satisfies the same quantization
condition $\int F_5\in 2\pi\Z$. Thus the supergravity
action for $F_5$ is fixed  to be
\begin{equation}
S_{F_5}=\frac1{4\pi} \int_{\AdS\times X} \cF_5 \wedge *\cF_5
\end{equation} 
where $\cF_5=F_{0,5}+F_{1,4}+F_{2,3}$.

Plugging \eqref{metricansatz} and \eqref{fluct}  into the ten-dimensional action, we obtain
\begin{equation}\label{tauij}
\tau_{IJ}=  \frac{N^2}{2\pi}
    \int_X (\omega_J\wedge *\omega_I+\frac{1}{16V^2}\contract{k_J}{k_I}\vol)
\end{equation}where the first and second terms come from the kinetic
terms for the five-form and the metric, respectively. This expression for $\tau_{IJ}$ agrees with the one presented in
\cite{currentcorrelators}.
Then, from \eqref{final}, we finally obtain \begin{equation}
c_{IJK}
=\frac{N^2}{2}
\int_X \omega_{\{I} \wedge \iota_{k_J} \omega_{K\}} \label{cijk}.
\end{equation}

\subsection{$a$ and the volume}
\def\Vol{\mathrm{Vol}}
Before moving to the explicit evaluation of $c_{IJK}$
for various Sasaki-Einstein manifolds,
let us determine the central charge $a$ from our  formula \eqref{cijk},
and check that it is inversely proportional to the volume.
In this subsection, we assume $X$ is not just an Einstein
manifold but also is Sasaki-Einstein.

Let $J$ be the K\"ahler form of the cone $C(X)$ over $X$,
and $e_r = r\partial_r$ the dilation on the cone direction.
Let $e$ be the one-form $\iota_{e_r} J$.
It endows $X$ with the structure of a contact manifold
so that $\vol_X=e\wedge J \wedge J/2$ and $de=2J$.
The Reeb vector is  $ i e_r$.

Since $X$ is now Sasaki-Einstein, the corresponding
CFT is $\N=1$ supersymmetric.
Let the R-symmetry in the superconformal algebra
be the linear combination $R^I Q_I$.
Then,  the central charge $a$ is given by \begin{equation}
a=\frac{9}{32} c_{IJK} R^I R^K R^K
= \frac {N^2} 2 \frac{27}{16} \int \omega_R\wedge \iota_{k_R}\omega_R\label{a}
\end{equation}where $\omega_R=R^I\omega_I$ and $k_R=R^I k_I$.
It is known through the work \cite{BHK} that
 $\omega_R$ is a multiple of $e\wedge J$.
We should normalize it so that $k_R$ is proportional
to the Reeb vector, and the holomorphic three-form $\Omega$ on $C(X)$ has
charge $2$ under $k_R$. Thus, we obtain
\begin{equation}
k_R = 2\pi \frac{2}{3}ie_r
\end{equation} because
$\Omega$ scales as $r^3$ and
the natural holomorphic one-form is $r e$.
The extra factor of $2\pi$ comes from our
convention $k_I=2\pi c^a_I k_a$
relating $k_I$ and the $k_a$ in the metric ansatz.

Thus, we have \begin{equation}
\omega_R= - \frac{\pi e\wedge J}{3V}
\end{equation} from \eqref{closeduptoisometry}.
Then eq \eqref{a} becomes \begin{equation}
a=\frac{N^2}2 \frac{27}{16}  \frac{4\pi^3}{27} \frac{\int e\wedge J\wedge J }{V^2}
=\frac{N^2}{4} \frac{\pi^3}{V},
\end{equation} which is precisely the relation established in \cite{HS,Gubser}.

\section{Properties of the supergravity formula}

\subsection{Giant Gravitons and  the normalization of $\omega_I$}\label{gg}

We have found so far the formula \eqref{cijk} for the CS coefficient $c_{IJK}$
given in terms of three-forms $\omega_I$ on the Einstein
manifold $X$.
The gauge field in the $\AdS$ space has these forms
as wavefunctions.
In order to compare the result to the field theory in four dimensions,
first we need to find the basis of the gauge fields so that
charged objects have integral charges with respect to these gauge fields.

Let us recall the situation in the compactification of the M-theory on a Calabi-Yau $Y$.
In that case, a massless gauge field arises from the M-theory three-form
with a harmonic two-form $\omega$ on $Y$ as the wavefunction,
and harmonic two-form naturally corresponds to $H^2(Y,\R)$.
M2-branes wrapped on a two-cycle $C$ in the Calabi-Yau give rise to
the charged particles in the noncompact dimensions,
and the charge is given by $\int_C \omega$.
Thus, 
$H^2(Y,\Z)\subset H^2(Y,\R)$ gives the integral basis we wanted.

Similarly in our case,
D3-branes wrapped on three-cycles in the Einstein manifold $X$
give rise to charged objects in the $\AdS$ side\footnote{The R-charge of the
wrapped D3-branes was studied in \cite{BHK}. The analysis of the R-charge and
the baryonic charges in the regular Sasaki-Einstein manifolds
was carried out in detail in \cite{HerzogWalcher}.}.
There are $b^3(X)$ homologically independent three-cycles.
We also have $\ell$ Kaluza-Klein angular momenta  associated to the
$\ell$ isometries.  For example, gravitons moving inside $X$ will
give charged objects from the $\AdS$ point of view.
In all, there are $d=b^3(X)+\ell$ types of charged objects
which match the number of the massless gauge fields.

Let us give a simple argument showing that ordinary homology of
3-cycles is not the correct mathematical object
to classify the charges of the  supersymmetric wrapped D3-branes.
For $S^5$ the homology is trivial but there are
giant gravitons. A less simple example comes from the $Y^{p,q}$
geometries (where the topology is simply $S^2 \times S^3$): there are various supersymmetric 3-cycles which are
homologically equivalent but have different volumes. D3-branes
wrapped on different cycles correspond to different operators in
the dual quiver gauge theory. These SUSY 3-cycles are invariant
under the $U(1)^l=U(1)^3$ isometries. The point is that we cannot
deform one such SUSY 3-cycle to another keeping it invariant under
the isometries. It is thus clear that we need some kind of
homology that keeps track also of the isometries, which
show up in $\AdS_5$ as Kaluza-Klein momenta.

Alert readers might be puzzled by now by the fact that the wavefunctions
$\omega_I$ are not closed in general.  Then the charge of a
wrapped D3-brane depends not only on its homology class, but also
on extra data, as expected also from the discussion in the
previous paragraph.
The Kaluza-Klein gauge fields coming from the
metric also enter the expansion of $F_5$, because in the expansion
\eqref{fluct} \begin{equation} \delta F_5= d(A^I \wedge N
\omega_I),
\end{equation} $A_I$ includes the gauge fields from the metric through \eqref{mixing}.
The non-closedness of $\omega_I$
allows a D3-brane wrapping a topologically trivial cycle $C$
to have non-zero coupling to $A^I$ given by \begin{equation}
N\int_C\omega_I.\label{Dbranecharge}
\end{equation}
For instance, if we consider Type IIB theory on
$S^5$ with $N$ units of  five-form flux and
we wrap a D3-brane on $S^3$ at the equator,
it will give rise to a soliton with $N$ unit of Kaluza-Klein momenta.
This is precisely the maximal giant gravitons
treated in \cite{invasion,susygoliath}.

For simplicity, let us restrict our attention to branes
which are not moving in the SE.  In order for them to be charge
eigenstates, their worldvolume should be invariant under
the isometry.
Let us introduce an equivalence relation such that
$C\sim C'$ if $C-C'=\partial B$ where $B$ is an invariant four-chain.
Then,  the coupling of the branes to the gauge fields $A^I$
depends only  on the equivalence class, because \begin{equation}
\int_{C}\omega_I-\int_{C'}\omega_I=\int_{\partial B}\omega_I
=\int_B d\omega_I =\int_B \iota_{k_I}\vol^\circ,
\end{equation} and  the integral of $\iota_{k}$ acting
on anything vanishes if the integration region $B$ is invariant under $k$.
It is because the integrand is zero when $k$ is degenerating on $B$ and
 the interior product kills the legs
along $B$ when $k$ does not degenerate on $B$.

Suppose  $X$ has $U(1)^\ell $ isometry
and the third Betti number is  $b^3$.
In the explicit examples we will treat in the following sections,
there are always  $d=\ell +b^3$ of independent invariant three-cycles,
although we could not find a general proof in the mathematical literature\footnote{
In \cite{Mikhailov,Beasley}, one can find
interesting discussions on the construction of the supersymmetric three-cycles
using the complex algebraic geometry of
 the cone over the Sasaki-Einstein manifolds.}.
Assuming this, D3-branes wrapping on invariant three-cycles
comprise a good basis of charged objects with respect to
the gauge fields $A^I$.  Let us denote the basis by $C^I$, ($I=1,\ldots,d$).
Then, \begin{equation}
\int_{C^I}\omega_J =  \delta^I_{J},\label{normalization}
\end{equation}
determines the dual basis for the wavefunctions of the gauge fields $A_I$.
Then a D3-brane wrapping the cycle $C^I$ has charge $N$ under $A_I$,
and charge $0$ for other gauge fields.

\subsection{Metric independence of $c_{IJK}$}

First we recall the situation for the M-theory on  Calabi-Yau 3-fold case.
There,  after the Kaluza-Klein reduction,
the five-dimensional Chern-Simons interaction $c_{IJK}$ of the massless
gauge fields $A^I$
is given by \begin{equation}
c_{IJK}\propto \int \omega_I\wedge \omega_J\wedge \omega_K\label{above}
\end{equation}where $\omega_I$ is the two-form on the Calabi-Yau
which appears in the Kaluza-Klein Ansatz for the M-theory three-form $C$,
\begin{equation}
\delta C=A^I\wedge \omega_I.
\end{equation} The masslessness of $A_I$ requires $\omega_I$  to be harmonic,
and explicitly finding the harmonic form is quite difficult.  Fortunately,
the formula above \eqref{above} is independent of the shift of
$\omega_I$ by exact forms.  It implies that
$c_{IJK}$ becomes independent of the metric.

Similarly, we found in sec. \ref{sugra} the form $\omega_I$ is co-closed
and `closed up to isometry' \eqref{closeduptoisometry}.
We show in this section that
$c_{IJK}$ and the normalization condition
do not change under the shift \begin{equation}
\omega_I\to \omega_I + d\alpha_I + \iota_{k_I}\beta
\end{equation}where $\alpha_I$ are two-forms, $\beta$
is a four-form, both of which are assumed to be invariant
under $U(1)^\ell$ action.

First we discuss the shift $\omega_I\to \omega_I + d\alpha_I$.
The normalization condition \eqref{normalization} is not affected.
The change in  $c_{IJK}$ is zero because
\begin{equation}
\delta c_{IJK}\propto  \int d\alpha_{\{I }\wedge  \iota_{k_{J}} \omega_{K\}}
=-\int \alpha_{\{I} \wedge \iota_{k_{J}} d\omega_{K\}}\\
=-\int \alpha_{\{I} \wedge \iota_{k_{J}}\iota_{k_{K\}}} \vol_X^\circ=0.
\end{equation}

Secondly, we  turn to the shift   $\omega_I\to \iota_{k_I}\beta$.
Here, we need to
shift all of the forms $\omega_I$  simultaneously using the same $\beta$.
It induces the change in $c_{IJK}$ by \begin{align}
\delta c_{IJK}=\int \iota_{k_{\{I}}\beta \wedge \iota_{k_{J}} \omega_{K\}} &=0.
\end{align} Hence it does not change the CS coefficient.
As for the normalization \eqref{normalization},
the cycles $C^I$ are assumed to be invariant under the  isometry.
Then we have $\int_{C^I} \iota_{k_J} \beta=0$, using the same argument
as before.

From the relation \eqref{closeduptoisometry},
the shift $\omega_I\to \iota_{k_I}\beta$ is accompanied by the shift
$\vol^\circ \to \vol^\circ - d\beta$.  It means that we are free to take
any five-form which integrates to one as $\vol^\circ$ in determining
$\omega_I$ through \eqref{closeduptoisometry}.
The equation \eqref{closeduptoisometry} fixes $\omega_I$ only
up to the addition of exact forms, which was shown not to affect
$c_{IJK}$ above.

Let us recapitulate the method to calculate $c_{IJK}$.
\begin{itemize}
\item
We first take any invariant
five-form $\vol^\circ$ which satisfies $\int \vol^\circ =1$.
\item
Then find $\omega_I$ with the normalization $\int_{C^J}\omega_I=\delta^J_I$, \eqref{normalization}.
\item
Next we define $k_I$ as the linear combination of $\ell$
isometries such that the condition $d\omega_I+\iota_{k_I}\vol^\circ =0$, \eqref{closeduptoisometry} is satisfied.
\item
Finally we plug these quantities to the formula \eqref{cijk} and evaluate.
\end{itemize}
The procedure does not require knowledge of the
Einstein metric on $X$. We would like to emphasize that
the Sasaki structure on $X$ is not necessary in the calculation of $c_{IJK}$
either.
The only ingredient is the action of $U(1)^\ell$ on $X$.
In this sense we claim that $c_{IJK}$ is a topological invariant
of the manifold with $U(1)^\ell$ action.

\section{Explicit Evaluation of the supergravity formula}
\subsection{Sasaki-Einstein manifolds with one $U(1)$ isometry}
We first treat the case where there is only one isometry $k$ on
the Sasaki-Einstein manifold $X$.
We take the period of $k$ to be $2\pi$.
Then, the isometry determines on $X$
an $S^1$ fibration \begin{equation}\begin{array}{ccc}
 S^1&\to& X  \\
 &&\downarrow\\
 && B
 \end{array}\end{equation} over a K\"ahler-Einstein base $B$.
Let the one-form $e$ be $e=g_{ij}k^i dx^j$.
Then, the Sasaki-Einstein condition implies that
the curvature of the circle bundle $de$ is equal to twice the
K\"ahler class $J$ of the base $B$, that is,\begin{equation}
de=2J.
\end{equation} We have $\vol^\circ\propto e\wedge J\wedge J$.
Then, an elementary calculation shows that
elements of $H^3(X)$ corresponds
to elements of $H^2(B)$ annihilated by $J\wedge $.
Thus, $b^3(X)=b^2(X)-1$.  Since we assumed
$\ell=1$, the number of the gauge field $d$ is
\begin{equation}
d=\ell + b^3(X)=b^2(B).
\end{equation}

Thus, we need to find $b^2(B)$ of three-cycles $C^I$ and
three-forms $\omega_I$ in $X$
which satisfy the constraint \eqref{closeduptoisometry} and  \eqref{normalization}.
To this end, take a basis of two-cycles $D^1,\ldots,D^d$ in $B$
and the dual basis of two-forms $\gamma_1,\ldots,\gamma_d$ on $B$
such that $\int_{D^I}\gamma_J=\delta^I_J$.
Let us take $C^I$ to be the three-cycle above $D^I$ in the fibration
and $\omega_I= (2\pi)^{-1}e\wedge \gamma_I$.
Then the normalization \eqref{normalization} is automatic, and
from $d\omega_I+\iota_{k_I}\vol^\circ =0$ \eqref{closeduptoisometry}, we have
\begin{equation}
k_I=  - 2(\int_B J\wedge \gamma_I ) k.
\end{equation}
Thus we obtain\begin{equation}
c_{IJK}=\frac{N^2}2 \int_B \frac{J}{\pi}\wedge \gamma_{\{I} \int_B \gamma_J\wedge\gamma_{K\}}
. \label{ell=1formula}
\end{equation}

\subsection{Higher del Pezzo surfaces}\label{delpezzo-gravityside}
Circle bundles over del Pezzo surfaces are prime examples of
five-dimensional Sasaki-Einstein manifolds, where
the $n$-th del Pezzo surface $dP_n$ for $n<9$ is $\CP^2$ blown up at generic $n$ points.
For $n=1,2,3$ they are toric, which will be treated in the next subsection.
In this subsection we evaluate \eqref{ell=1formula} for del Pezzo surfaces with $n\ge 4$,
which have only one isometry which rotates the circle fiber.
We  compare the result with the field theory result  in section \ref{delpezzo-CFTside}.

Let us take $\gamma_0$ as the two-form dual to the base $\CP^2$,
and $\gamma_i$, $i=1,\ldots,n$ be the two-forms dual to the $i$-th
exceptional cycle.  The intersection paring is Lorentzian, i.e. \begin{equation}
\int_{\dP_n} \gamma_I \wedge \gamma_J = \diag(+1,-1,\ldots,-1).
\end{equation} where $I,J=0,1,\ldots, n$.
The K\"ahler form $J$ is chosen to be equal to
negative of the Chern class of the anti-canonical bundle,\begin{equation}
J=\frac{\pi}{3} (3\gamma_0-\sum_{i=1}^n \gamma_i).
\end{equation} The area of the $\dP_n$ is $\int_{\dP_n} J\wedge J / 2 =\pi^2 (9-n)/18$.
Formula \eqref{ell=1formula} can be conveniently packed in the cubic polynomial
\begin{equation}
P_n(a_0,a_1,\ldots,a_n)\equiv  c_{IJK}a^Ia^Ja^K
= 3 N^2 \int_{\dP_n} \frac{J}{\pi}\wedge \gamma  \int_{\dP_n} \gamma \wedge\gamma
\end{equation} by
introducing indeterminate variables $a^I$, $I=0,\ldots,n$ and  $\gamma\equiv\gamma_I a^I$.
It can be easily evaluated to be \begin{equation}
P_n(a^I)=N^2 \left(3a^0 + \sum_i a^i \right)\left((a^0)^2 - \sum_i (a^i)^2 \right).
\end{equation}
An  obvious consequence is that we have \begin{equation}
P_n (a^0,a^1,\ldots, a^n)= P_{n+1}(a^0,a^1,\cdots,a^n,a^{n+1}=0).
\label{invariance-delpezzo}
\end{equation} We will see the physical mechanism behind this result
in later sections.

\subsection{Toric Sasaki-Einstein manifolds}
\begin{figure}
\centerline{\includegraphics[height=6cm]{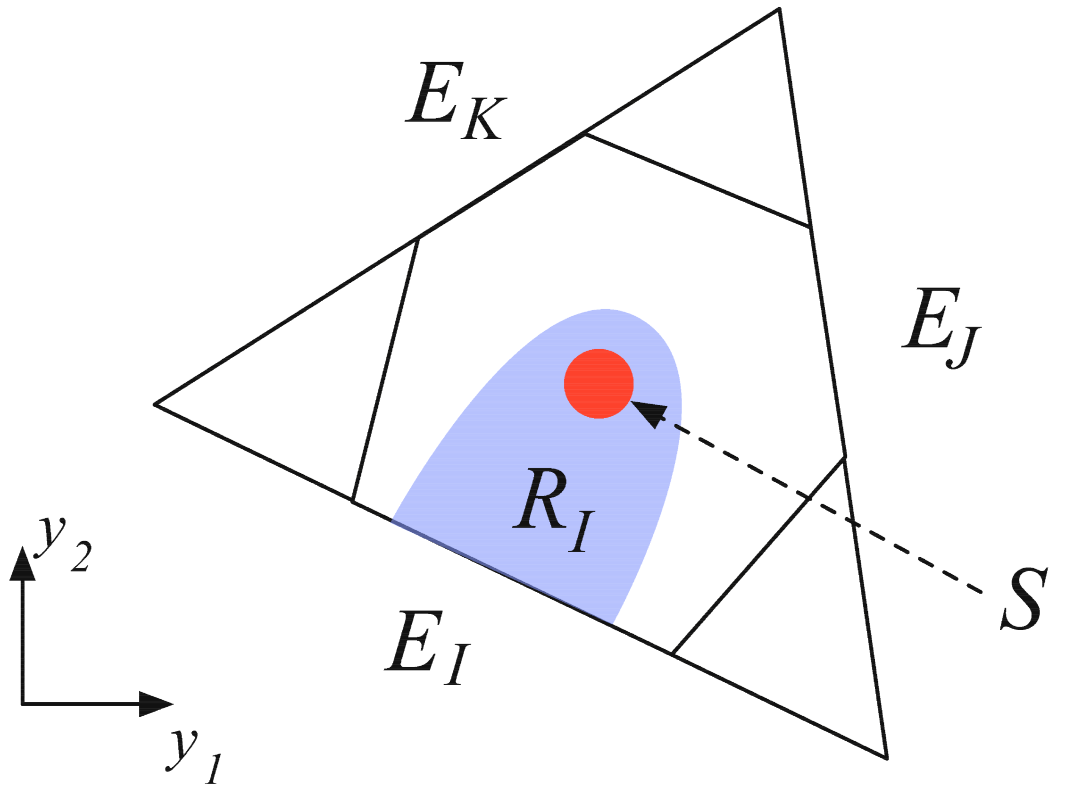}}
\caption{ Construction of $\omega_I$. The polygon designates the
image of the moment map. The red blob $S$ is
the support of $\cF$ and the blue region $R_I$ is the
support of $\cA_I$.  \label{figure}}
\end{figure}

We would like to move on to the case where there are
three isometries in the Sasaki-Einstein manifold $X$, i.e. $\ell=3$.
In that case, the Calabi-Yau cone over $X$ is toric,
thus $X$ is called  a toric Sasaki-Einstein manifold.
Let us describe $X$ as a $T^3$ fibration
over a two-dimensional $d$-gon $B$,
where the coordinates of $T^3$  are $\theta_{1,2,3}$
and those of the base are $y^{1,2}$.
We take the periodicity of $\theta_i$ to be $1$.
Denote the edges by $E^I$, $I=1,\ldots,d$,
the 3-cycles above them by $C^I$.
It is known that $H^3(X)=d-3$ so that
the number of the edges is precisely the number of gauge fields
which we obtain by compactifying Type IIB string on $X$.
Let $k_I=k_{iI}\partial/\partial\theta_i$ be the degenerating
Killing vector at $C^I$, see figure \ref{figure}.

We will see shortly that the
calculation of $c_{IJK}$ only depends on $k_{I,J,K}$ and
not on the other $k_{L\ne I,J,K}$ or the number of the edges.
From now on, all the forms are assumed to depend only on $y^{1,2}$.

Firstly, take a two-form $\cF$ on the base $B$ supported on a region $S$
with $\int \cF=1$. $S$ is marked with red in the figure \ref{figure}.
Choose \begin{equation}
\vol^\circ=\cF \wedge d\theta_1\wedge d\theta_2\wedge d\theta_3
\end{equation}
as the normalized volume form.

Secondly, for each edge $E_I$, draw a region $R_I$ which contains $S$
and touches only with $E_J$ with $J=I$ (cf. fig. \ref{figure}).
Choose the one-form $\cA_I$ on the base $B$
which is non-zero only in $R_I$ such that $d\cA_I=\cF$.
Notice that $\int_{E^I}\cA_J=\delta^I_J$,
since $\cA_J$ is only nonzero on $R_J$ and
$\sum_J\int_{E_J} \cA_I=\int_B \cF=1$.

We need to ensure furthermore \footnote{
The construction of the forms $\cA_I$ can be done as follows:
Let the $x$-axis be along the edge $E_I$, the $y$-axis be perpendicular to it,
and the region $R_I$ be given by $0\le y\le a(x)$. Denote
$\cF=F(x,y) dx\wedge dy$. Then  $\cA_I=dx  \int_y^{a(x)} F(x,y)dy$
satisfies the required properties. It can be done similarly
for other more complicated shape of $R_I$.}
that $\cA_I$ has only components
parallel to the edge $E_I$.
Then \begin{equation}
\omega_I\equiv -\cA_I\wedge \iota_{k_I}d\theta_1\wedge d\theta_2\wedge d\theta_3
\qquad (\text{no summation on $I$})\label{explicitchoicefortoric}
\end{equation}
is a well-behaved form on $X$,
since the existence of $\iota_{k_I}$ guarantees that $\omega_I$ is
regular near $E_I$, and the fact $A_I$ vanishes outside the blue region
guarantees $\omega_I$ is regular near $E_{J\ne I}$.
It also satisfies the constraint \eqref{closeduptoisometry} and \eqref{normalization}
almost by construction.

Now we can clearly see that the forms $\omega_{I,J,K}$ can be taken to be  the same
irrespectively of, for example,
whether we are calculating $c_{IJK}$ for the hexagon inside
or the triangle outside in the figure.
Thus, $c_{IJK}$ depends only on
$k_{I,J,K}$ and not at all on $k_{L\ne I,J,K}$.
It is even independent of the number of the edges, i.e.\begin{equation}
c_{IJK}=f(k_I,k_J,k_K).\label{independence}
\end{equation}

First of all, if two of $k_{I,J,K}$ are equal, then $f$ is obviously zero
because the integrand is zero.
Next, let us consider the case when they are all different.
We can assume the base $B$ is a triangle without loss of generality.
We will show that $X$ is an orbifold of $S^5$,
which allows us to obtain $c_{IJK}$.

Take the universal cover $U$ of $X$,
that is, remove the periodicity $\theta_i\sim\theta_i+1$.
$X$ can be obtained by
dividing $S^5$ with  the lattice $N$ generated by $(\theta_1,\theta_2,\theta_3)=(1,0,0)$,
$(0,1,0)$ and $(0,0,1)$.
Instead, consider a manifold $Y$ by dividing
$U$ by the lattice $L$ generated by $k_I$, $k_J$ and $k_K$.
Along the edges of $B$, precisely the direction
$k_{I,J,K}$ degenerates. Thus we have shown that $Y$ is topologically an $S^5$,
and $X=S^5/\Gamma$ where $\Gamma$ is the finite group $L/N$.
The order of $\Gamma$ is \begin{equation}
\#\Gamma=\left| \det (k_I,k_J,k_K) \right |.
\end{equation}

\begin{figure}
\begin{center}\includegraphics[height=6cm]{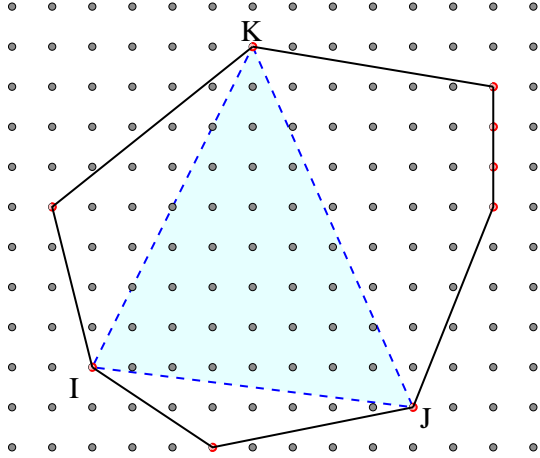}\end{center}
 \caption{Pictorial representation of the toric formula $c_{IJK}=\frac{N^2}{2} \left| \det(k_I, k_J,k_K)
 \right|$.
\label{areat}}
\end{figure}

Let us denote the corresponding quantities on $S^5$ by adding tildes
and the projection map by $i:S^5\to S^5/\Gamma=X$,
we find \begin{equation}
i^*\omega_I = (\#\Gamma) \tilde\omega_I, \quad i^*\vol^\circ=(\#\Gamma)\,\widetilde\vol^\circ,
\quad \text{and}\quad  i^*k_I=\tilde k_I.
\end{equation} Then \begin{equation}
\int_{S^5/\Gamma}  \omega_{\{I} \wedge \iota_{k_J} \omega_{K\}}
=(\#\Gamma)^{-1}
\int_{S^5}  i^*\omega_{\{I} \wedge \iota_{k_J} i^*\omega_{K\}}
= \#\Gamma
\int_{S^5}  \tilde\omega_{\{I} \wedge \iota_{\tilde k_J} \tilde\omega_{K\}}
\end{equation}that is, $c_{IJK}$ is $\#\Gamma$ times that of $S^5$.
Finally, for $S^5$, one can do the explicit calculation to find
$c_{IJK}=N^2/2$.
Thus we obtain the formula \begin{equation}
c_{IJK}=\frac{N^2}{2} \left| \det(k_I, k_J,k_K) \right|,\label{sugraresult}
\end{equation}which is proportional to the area of the triangle
inside the toric diagram, see figure \ref{areat}.

\section{Field Theory Analysis}
From AdS/CFT duality,
there are global  symmetries $Q_I$ and their currents $J_I$ on the boundary
corresponding to the gauge fields $A^I$ in the bulk
with the boundary coupling \begin{equation}
\int d^4x A^I J_I.
\end{equation} Thus,
the Chern-Simons interaction $c_{IJK}$
in the five-dimensional action
induces the triangle anomaly on the CFT side \cite{W}.
The numerical coefficient in \eqref{5dlagrangian} is chosen such
that\begin{equation}
c_{IJK}=\tr ( Q_I Q_J Q_K )
\end{equation}is satisfied.
We obtained a concrete supergravity formulae for $c_{IJK}$
in the previous sections.
We also know the corresponding quiver theories which
flow to CFTs in the IR through recent developments.
We will see that the triangle anomaly calculated in the
quiver side completely agrees with the supergravity calculation.

\subsection{Cubic anomalies from field theories in the toric case}\label{ft}
In this section we compute all the cubic 't Hooft anomalies in the
case of gauge theories dual to toric Sasaki-Einstein manifolds. In
order to perform this computation, we first have to know the
structure of the quiver theory.
Although we will summarize below only the facts that we need later,
the method  of obtaining the quiver gauge theory from
the toric data and vice versa
is a beautiful subject in itself.
It has been known for some time
that it can be done in principle algorithmically,
but the method was unwieldy and required extensive calculation.
Now various works, \cite{dimers,dimers1,geo,LpqQuiver0,LpqQuiver,BZ,Butti:2005ps,HV2005,KFHV},
give a technique to obtain the quiver theory
in a much more streamlined way by the so-called dimer methods.
They accomplished the most difficult parts at the same time, namely
the determination of the superpotential of the quiver theory. 
We would like the reader to refer to the
works cited above for these developments.

We will use the following properties of the quiver gauge theories
dual to a toric diagram:
\begin{enumerate}
 \item The gauge group is $SU(N)^{\mathcal{A}}$, where $\mathcal{A}$
 is twice the area of the toric diagram.
 \item The bifundamental chiral superfields
can be grouped in $d(d-1)/2$ sets, which we can call
$\mathcal{B}_{i j}$, where $i$ and $j$ label two external
 $(p,q)$-legs. In each set $\mathcal{B}_{i j}$ there are
 \begin{equation}\label{eqdet}
 |\mathcal{B}_{i j}| = p_i \, q_j - p_j \, q_i
 \end{equation}
of bifundamental fields, where $(p_i, q_i)$ is the $i$-th
external $(p,q)$-leg.
\item All the fields belonging to the same set
 $\mathcal{B}_{ij}$ have, under the global symmetry $U(1)^d$,
 the same charges $Q^{ij}_I$.
\end{enumerate}
The full group of global symmetries, as we saw, is
 \begin{equation}
U(1)^d = U(1)^3_F \times U(1)_B^{d-3}
 \end{equation}
if the toric diagram has $d$ points on the boundary. 

Before proceeding let us comment on what is known about the
validity of the various properties. Property $1$ is a well
established fact. The total number of gauge groups is equal to the
total number of compact cycles ($0$-, $2$- and $4$-cycles) in the
completely resolved Calabi-Yau. Since there is no odd-homology,
this number is the Euler number of the resolved non compact
Calabi-Yau, which is, in turn, given by twice the area of the
toric diagram. Properties $2$ and $3$ were proposed in \cite{geo},
under the name of ``folded quiver''. Property $2$\footnote{
We expect there is always at least one toric phase where the
number of the fields is precisely given by the determinant
(\ref{eqdet}). This is known to be the case for the set of
theories $Y^{p,q}$ and $L^{p,q|r}$. For the $Y^{p,q}$'s all toric
phase have been classified \cite{BHK}, and in some phases, with so
called double impurities, property $2$ does not hold as stated. In
these cases there are additional pairs of fields with opposite
charges} was shown for toric del Pezzo surfaces in
\cite{HananyIqbal}, and there is by now a lot of evidence for it, for
instance the exact quiver gauge theories are known for
$Y^{p,q}$/$L^{p,q|r}$ and they satisfy property $2$. We expect it
to be possible to give a general proof studying intersection
numbers of compact three-cycles in the mirror Calabi-Yau, as was
conjectured in \cite{geo} on the base of \cite{HananyIqbal}. For
recent work see \cite{HV2005, KFHV, Butti:2005ps}. In particular using
the procedure devised in \cite{HV2005} it is possible to derive
formula (\ref{eqdet}) from the counting of the intersection of
$(p,q)$-legs when drawn in the planar torus (again, consult
\cite{HV2005} for details). Let us stress
that the properties $1$ and $2$ are inherently topological, in the
sense that the former depends only on the topology of the
Calabi-Yau
 and the latter that of its mirror. Property $3$ instead goes slightly
beyond purely topological properties, for instance the existence
of three $U(1)$ flavor symmetries is related to isometries of the
Calabi-Yau metric. Let us notice also that in \cite{HananyIqbal} a different
interpretation of (\ref{eqdet}) was given, and we now know that the correct
interpretation is in terms of property $3$.

Very strong evidence for the validity of all the three properties
listed above was given in the work of Butti and Zaffaroni
\cite{BZ,Butti:2005ps}, where it was shown that the field theory
computation of the cubic 't Hooft anomaly $c_{RRR}$ matches
precisely the geometric results for the volumes of the
Sasaki-Einstein, as expected from $\AdS/\CFT$ correspondence. The
volumes on the gravity side can be computed using the results of
Martelli, Sparks and Yau \cite{MSY}, which enables to compute the
volumes just in terms of toric data. We will show that \emph{all
cubic 't Hooft anomalies $c_{IJK}$ match with the Chern-Simons
coefficients as computed from gravity}.

As an aside, let us note that, beyond 't Hooft anomalies, using
the ``folded quiver'' picture, one can readily compute the scaling
dimension of dibaryon operators and succesfully match with string theory. This gives additional evidence
for the validity of properties $1, 2$ and $3$. Also the topology
of some SUSY three-cycle can be matched with this picture
\cite{LpqQuiver0}.

\begin{figure}
\begin{center}\includegraphics[height=7cm]{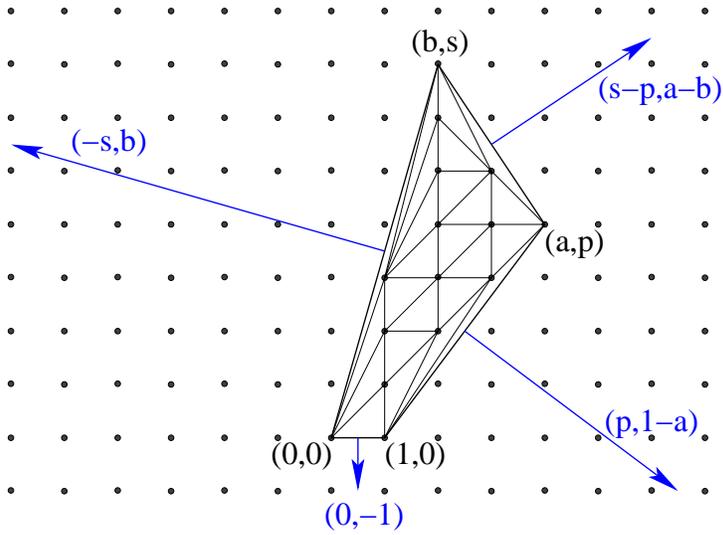}\end{center}
 \caption{A generic toric diagram with
 four corners, i.e. a generic $L^{p,q|r}$, and the associated $(p,q)$-web. We have $s=
 p+q-r$. The integers $a$ and $b$ are such that $as-bp=q$.
\label{fig0}}
\end{figure}

\begin{figure}
\begin{center}\includegraphics[height=7cm]{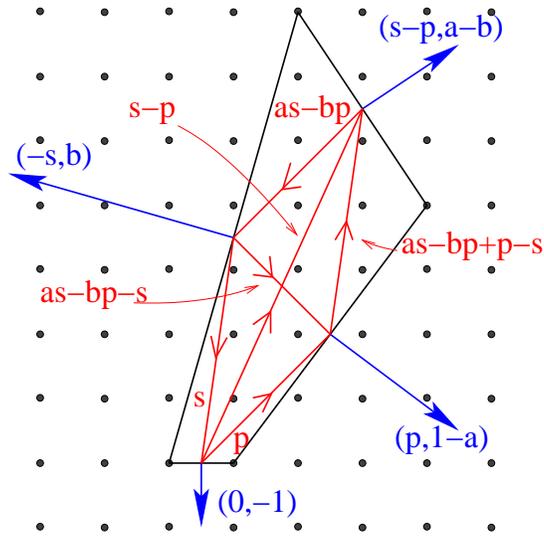}\end{center}
 \caption{An example of ``folded quiver.'' From a generic toric diagram with
 four nodes we can immediately compute the multiplicities of $6$ sets of bifundamental
 fields.
\label{fig}}
\end{figure}

In order to compute the full set of cubic 't Hooft anomalies we
need to identify the $d$ $U(1)$ global symmetries. We will take
all the $d$ symmetries to be $R$-symmetries (taking linear
combinations it is obvious how to obtain $d-1$ ordinary $U(1)$
symmetries).
There is a natural way to associate a $U(1)$ symmetry to every
external node in the toric diagram: the charge of a field under the
$i$-th symmetry is one if  the $i$-th node on the right
of the arrow corresponding to the field
in the folded quiver diagram, zero otherwise. For
instance external fields in the folded quiver diagram are charged
only under one $U(1)$ symmetry. 
In
this way all chiral superfields have charges $0$ or $1$ under the
$U(1)^d$ global symmetry.
The superpotential corresponds to closed loops of the folded quiver.
Thus, its charge under the $i$-th $U(1)$ symmetry
is $1$. It implies that the commutation relation between
$I$-th $U(1)$ charge $Q_I$ and the supercharge $Q_\alpha$ is
$[Q_I,Q_\alpha]=-Q_\alpha/2$.
This in turn means that 
the gauginos have thus charge one half and
their contribution to cubic anomalies is always $\mathcal{A} N^2/8$.
 Then, the  fermionic component of the bifundamental  superfields
have thus charge $-1/2$ or $1/2$. We thus see that in this way all the
charges are half integral, and every bifundamental field contributes
$\pm N^2/8$ to the cubic anomalies. The point is that this basis is
precisely the field theory dual of the basis considered in the
previous subsection. Indeed, the dibaryon constructed from the field
in $\f_{I,I+1}$ has the charge $\delta_{IJ} N$ under the symmetry $Q_J$,
which precisely matches the charge of the D3-brane
which wraps the cycle $C_I$, see \eqref{normalization}.

Let us report in detail the results for the case of toric diagram
with $4$ corners. The charges are given by table \ref{charges1}.

\begin{table}[!h]
\def\mm{\phantom{-}}
\begin{center}
$$\begin{array}{|c||c||c|c|c|c||c|c|c|c|} \hline
\ \mbox{Field}\ & \ \mbox{Number} \ & Q_1 &\ Q_2 \ &\ Q_3 &\ Q_4
&\ Q^F_1 &\ Q^F_2 \ &\ Q^F_3 &\ Q^F_4  \\ \hline\hline
 \f_{12} & p        & 1 & 0 & 0 & 0 & \mm1/2 & -1/2 & -1/2 & -1/2   \\ \hline
 \f_{23} & r        & 0 & 1 & 0 & 0 &  -1/2 & \mm 1/2 & -1/2 & -1/2    \\ \hline
 \f_{34} & q        & 0 & 0 & 1 & 0 & -1/2 & -1/2 &  \mm1/2 & -1/2   \\ \hline
 \f_{41} & p + q -r & 0 & 0 & 0 & 1 & -1/2 & -1/2 & -1/2 &  \mm1/2   \\ \hline
 \f_{13} & q - r    & 1 & 1 & 0 & 0  & \mm 1/2 &  \mm1/2 & -1/2 & -1/2   \\ \hline
 \f_{42} & r - p    & 1 & 0 & 0 & 1  &  \mm1/2 & -1/2 & -1/2 &  \mm1/2   \\ \hline
\text{ Gauge}   & p + q    & 0 & 0 & 0 & 0    &   \mm1/2 &  \mm1/2 & \mm 1/2 &  \mm1/2   \\ \hline
\end{array} $$
\caption{Charge assignments for the basic superfields in the case
of toric diagrams with four corners.} \label{charges1}
\end{center}
\end{table}

It is straightforward to check that the linear 't Hooft anomalies
vanish, i.e. $\tr (Q_j) = 0$. This has to be the case for any
superconformal quiver \cite{IW2003, BH2004}. A general proof of
the vanishing of linear anomalies using the folded quiver picture
was given in \cite{BZ}. Since $(Q^F_i)^2 = 1$, $\tr (Q_j) = 0$
also implies that
 \begin{equation}
 \tr (Q_i^2 Q_j) = \tr (Q_j) = 0
 \end{equation}

The remaining cubic 't Hooft anomalies (recall they are completely
symmetric) are easily computed to be
 \begin{eqnarray}
\tr(Q_1 Q_2 Q_3) & = &\,  N^2 \, r\,/2 \\
 \tr(Q_2 Q_3 Q_4) & = &\,  N^2 \, q \,/2\\
 \tr(Q_3 Q_4 Q_1) & = & N^2 (p + q - r)\,/2 \\
 \tr(Q_4 Q_1 Q_2) & = &\,  N^2 \,\, p\,/2
 \end{eqnarray}
It is now straightforward to check that these are proportional to the
area of the triangles  \begin{equation}
|\det(k_I,k_J,k_K)|
\end{equation} spanned by the corners of the toric diagram
of figure \ref{fig0} or \ref{fig}.
Thus we have shown that, for a toric diagram with four edges,
the cubic anomaly $c_{IJK}$ is given by \begin{equation}
c_{IJK}=\frac{N^2}2  |\det(k_I,k_J,k_K)|,
\end{equation} which agrees with  the supergravity result \eqref{sugraresult}.

This nice result can be proven
for a generic toric diagram with arbitrary number of edges,
by an easy mathematical induction.
We leave the details in the Appendix \ref{generaltoric}.

\subsection{del Pezzo surfaces}\label{delpezzo-CFTside}
Now we want to discuss the gauge theories corresponding to the
complex cones over smooth K\"ahler-Einstein surfaces, \emph{i.e.}
del Pezzo surfaces $dP_n$ for $3 \leq n \leq 8$. The quivers were
constructed in \cite{FHH2000} for toric del Pezzo surfaces
($dP_1$, $dP_2$ and $dP_3$), and in
\cite{HananyIqbal,unhiggsingdelpezzo} for the non toric ones, i.e.
$dP_n$ with $4 \leq n \leq 8$. The generic superpotential for
$dP_5$ and $dP_6$ was derived in \cite{WIJ}, for $dP_7$ and $dP_8$
the explicit, generic, superpotential is still not known. In
\cite{IW2003,hiddenexceptionalsym}, all the baryonic and $R$
charges are explicitly listed for $dP_n$ up to $n=6$. It is simple
to compute, using these data, the cubic 't Hooft anomalies and to
match with our geometrical findings in
sec.~\ref{delpezzo-gravityside}.

In \cite{HerzogWalcher}, the $R$- and baryonic charges of the
dibaryons were analyzed through the framework of the exceptional
collections on the del Pezzo surfaces. In particular, it was shown
that the triangle anomalies among the R-symmetry and two baryonic
symmetries, $\tr (R B_i B_j)$, are proportional to the
intersection form of the two-cycles which are perpendicular to the
K\"ahler class of the surface. It is easy to check that our
formula in sec.~\ref{delpezzo-gravityside} naturally reproduces
the result of \cite{HerzogWalcher}.

\section{Rolling down among Sasaki-Einstein vacua}
The triangle anomalies in the CFT side
and the Chern-Simons coefficients of the gravity side
showed a remarkable behavior.
Namely, for quiver theories for
toric Sasaki-Einstein manifolds,
the coefficient $c_{IJK}$ is determined solely by the
toric data $k_{I,J,K}$ and is independent of other $k_{L}$ for
$L\ne I,J,K$ \eqref{independence}.
We would like to give a heuristic
physical interpretation of this fact.
The same consideration can be applied to the del Pezzo  cases,
and its manifestation is \eqref{invariance-delpezzo}.
We concentrate on the toric cases below.

Consider a toric Sasaki-Einstein $X$ whose
dual toric diagram has $d$ edges.  Each edge $E_I$ naturally
corresponds to a global symmetry $Q_I$ in the quiver theory.
There are bifundamental fields $\Phi^I$ with charge $\delta_J^I$
under $Q_J$.  Then, we can form a dibaryon operator \begin{equation}
B^I=\epsilon_{i_1i_2\ldots i_N}\epsilon^{j_1j_2\ldots j_N}
\Phi^I{}^{i_1}_{j_1}\Phi^I{}^{i_2}_{j_2}\cdots\Phi^I{}^{i_N}_{j_N}.
\end{equation} It has the charge $N \delta_J^I$ under $Q_J$,
which is precisely the charge \eqref{Dbranecharge} of a D3-brane
wrapping the three-cycle determined by $E_I$.

\begin{figure}
\centerline{\includegraphics[height=3cm]{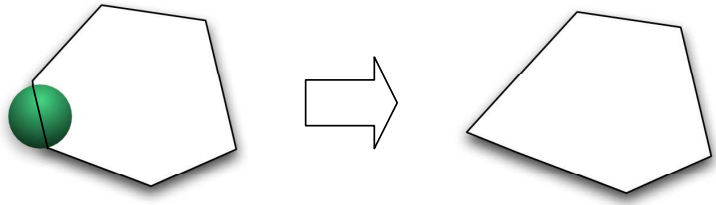}}
\caption{Schematic depiction of the dibaryon condensation.
Each edge corresponds to a three-cycle in the toric Sasaki-Einstein
around which  D3-branes can be wrapped.
Higgsing with the corresponding dibaryon operator in the quiver
CFT eliminates that edge.
\label{condensationfigure}}
\end{figure}

Now, let us give a vacuum expectation value (vev)
 to $B_I$. Since $B_I$ is charged only
with respect to $Q_I$ and not to $Q_{J\ne I}$,
the theory flow to a theory with $d-1$ global symmetries.
On the gravity side, the Higgsing means that there is
an infinite number of D3-branes wrapping around $C_I$,
which presumably shrinks it  just as in the blackhole condensation \cite{BHcond},
see figure \ref{condensationfigure}.
It is the blowdown of the toric divisor corresponding to $E_I$
on the Calabi-Yau cone over $X$. This procedure was used
in the determination of the del Pezzo quiver in \cite{unhiggsingdelpezzo}.

Recall that the same triangle anomaly can be calculated either in
the ultraviolet or in the infrared.
Thus, the triangle anomaly $c_{JKL}$ among the global symmetries
other than $Q_I$ is the same before and after the Higgsing.
Since the Higgsing eliminates
the edge $E_I$, this means that $c_{JKL}$ is independent of $k_I$.
One can repeat the flow many times
and we can reduce the toric diagram to  a triangle,
which is  an orbifold of $\mathcal{N}=4$ $SU(N)$
super Yang-Mills theory.

Let us consider the behavior of the central charge $a$ along the flow.
Consider a flow from the UV quiver theory to the IR quiver  theory
triggered by giving a vev to $B_I$.
The IR theory contains also a free chiral scalar
field which represents the fluctuation of the vev of $B_I$.
Its contribution to $a$ is of order $1/N^2$ compared to the
contribution from the interacting part, so we can neglect them henceforth.
Then, from the invariance of $c_{IJK}$ along the flow \eqref{independence},
the central charge $a$ in the IR theory can be obtained by maximizing the
same function as that for the UV theory in a smaller region.
Thus, $a$ will presumably decrease, with the usual caveat on the
fact that the trial function attains the maximum only locally.

Let us compare the process we saw in this section with the rolling among
Calabi-Yau vacua \cite{rolling}. There, theories
on various topologically-distinct Calabi-Yau manifolds
are connected by adiabatically changing the moduli.
Here, theories on various topologically-distinct
Sasaki-Einstein manifolds are connected by
the renormalization-group flow
induced by the Higgsing of the dibaryons.
Both have the same number of supercharges,
and both can be understood as the Higgsing.
Thus, we suggest to dub the phenomenon we found as the ``rolling among
Sasaki-Einstein vacua,'' although the rolling is
unidirectional.

More detailed analysis of the rolling is clearly necessary
and will be interesting.
We hope to revisit this problem in the future.

\section{Conclusion}
In this paper we explored  a particular aspect of the
$\AdS_5/\CFT_4$ correspondence. Namely, the matching between the
Chern-Simons interaction in the five-dimensional  bulk
and the triangle anomaly in the four-dimensional boundary.
More precisely,
we derived a formula for the Chern-Simons interactions
in terms of three-forms
in the Einstein manifold used in the compactification,
and we also evaluated the formula  for
the circle bundles over the del Pezzo surfaces
and for the toric Sasaki-Einstein manifolds.
Furthermore, we successfully matched the resulting expression
to the triangle anomaly  from the dual field theory.
Condensation of dibaryons was crucial in the physical understanding
of the calculation of the triangle anomaly in both sides of the duality.

We also found that the charges of the D-branes wrapping
various three-cycles in the Sasaki-Einstein
naturally and nontrivially combine
the angular momenta along the
isometry directions and the baryonic charges.

There are several open problems that we would like to point out.
One possible direction of further research is to extend the
determination of the lowest-derivatives terms in the $\AdS$ theory
and  to check the very  special geometry of the vector multiplet scalars.
Another direction will be the study of a more thorough understanding
of the charges of D-branes wrapping inside
the Sasaki-Einstein manifolds.
The new ingredients came in mostly from the fact that the manifold
comes with a group action. We made some comments
in the appendix \ref{more}.
Finally, the physics of the rolling among the Sasaki-Einstein vacua
should be studied more thoroughly.
We hope to revisit these
problems in the future.

\section*{Acknowledgement}
The authors would like to thank the Kavli Institute for
Theoretical Physics and the organizers of the workshop
``Mathematical Structure of String Theory'' held there for the
hospitality during the entire course of this work.  They would like to
thank the participants of the workshop for stimulating
discussions. Y.T. would like especially to thank Hirosi Ooguri for
fruitful discussion. They also would like to express gratitude to
Alastair King for discussions on mathematical aspects of the work.
S.B. is grateful to Agostino Butti and Alberto Zaffaroni for
useful discussions. L.A.P.Z. is partially supported by Department
of Energy under grant DE-FG02-95ER40899. Research of Y.T. is
supported by the JSPS predoctoral fellowship. The work is
partially supported by the National Science Foundation under Grant
No. PHY99-07949. The work of S.B. is supported by \emph{Fondazione
"Angelo della Riccia"}, Firenze, Italy.

\begin{appendix}

\section{Details of supergravity reduction}\label{details}
Our goal in this section is to perform a compactification of a
ten-dimensional solution of IIB supergravity to five dimensions.
It is worth stressing that we are not attempting a reduction to a
five-dimensional theory.  In fact, there is an extensive
literature on supergravity reduction on positively curved
symmetric manifolds. For example, there are some constructions of
full consistent non-linear Ansatz for the reduction on the spheres
\cite{diana}. Other interesting truncations are presented in
\cite{reduction} and references therein. In this subsection we
carry out the compactification of Type IIB theory on generic
5-dimensional Einstein manifolds. As such, we are forced to the
perturbative analysis and will not pursue full non-linear
reduction in this paper. Indeed, it is known that consistent
reductions are possible only for a restricted set of manifolds
\cite{consistency}.

Consider Type IIB theory compactified on an Einstein 5-manifold
$X$ to have a five-dimensional theory on $\AdS_5$. Let the
coordinates of $X$ and $\AdS$ be $y^i$ and $x^\mu$, and their
f\"unfbeine be $e^i$ and $f^\mu$, respectively.

Since the action of the self-dual five-form in ten dimensions is rather subtle,
we carry out the Kaluza-Klein analysis at the level of equation of motion.
Let us explain the main technical point before going into the details.
Schematically,
one first expands the fluctuation  using the harmonics of the internal
manifold $X$, \begin{equation}
\phi(x,y)=\phi_0(x,y)  +\delta \phi^{(i)}(x) \psi_{(i)}(y)+\cdots,
\end{equation}  so that $\delta\phi^{(i)}$ are the mass eigenstates.
Then, one can identify the cubic couplings such as the CS coefficient
by finding the equation of motion of $\delta\phi^{(i)}$ in the form \begin{equation}
( D-m^2 )\delta \phi^{(i)}= C^{(i)}_{(j)(k)} \delta\phi^{(i)}\delta\phi^{(j)}+\cdots.
\end{equation}
If one is only interested in obtaining certain parts of the cubic coupling,
one can set to zero any fluctuation which does not multiply the couplings.
It does not change the results, and at the same time
it  greatly reduces the calculational burden.

Another technical difficulty lies in maintaining
the self-duality of the Ansatz for the five-form.
Suppose $X$ has $\ell$  $U(1)$ isometries $k_a^i$, $a=1,\ldots,\ell$
with period  $2\pi$.
The ansatz for the metric is the usual one,\begin{equation}
ds^2_X=\sum_i (e^i+k_a^iA^a)^2, \label{metricansatzApp}
\end{equation}where $e^i$ are the f\"unfbein forms of the Einstein manifold
and $A^a=A^a_\mu dx^\mu$ are one-forms on $\AdS_5$.

Let us abbreviate $\hat e^i =e^i+k_a^i A^a$.
Then, the Hodge star exchanges \begin{equation}
f^1,\ldots,f^5\quad\longleftrightarrow\quad \hat e^1,\ldots,\hat e^5.
\end{equation}
Thus, one can anticipate that
the introduction of  the following $\hat{\phantom{M}}$
operation on differential forms of $X$
defined by replacing $e$ by $\hat e$, \begin{equation}
\alpha^{(p)}=\alpha_{i_1\cdots i_p} e^{i_1}\cdots e^{i_p}
\mapsto \hat \alpha^{(p)}\equiv \alpha_{i_1\cdots i_p} \hat e^{i_1}\cdots \hat e^{i_p},
\end{equation} greatly helps in maintaining the self-duality
of the Ansatz for $F_5$.

The following two formulae are useful in calculation.
First is a formula
for the $\hat{\phantom{o}}$ operation using interior products:
\begin{equation}
\hat \alpha = \alpha + A^a\wedge\iota_{k_a}\alpha
 + \frac12A^a\wedge A^b\wedge \iota_{k_b}\iota_{k_a}\alpha
 + \cdots.
\end{equation}
Another is $*(\alpha^{(5-p)}\wedge\beta^{(p)})= (-)^p (*\alpha)\wedge*\beta$
where the number in the parentheses in the superscript denotes the degree of the forms.

Let us carry out what we have just outlined.
The equations of motion  and the Bianchi identity in
Type IIB supergravity is:  \begin{align}
R_{\mu\nu}&= \frac{c}{24} F_{\mu....}F_{\nu}{}^{....},\label{metriceq}\\
F&=*F,\\
dF&=0
\end{align}where $R_{\mu\nu}$ is the Ricci curvature of the ten-dimensional metric
and $F$  is the self-dual five-form field strength.
The right hand side of \eqref{metriceq} should be contracted in a suitable way.
$c$ is a convention dependent constant.
We set other form fields and fermions to zero, and dilaton to constant.
In the following, we use the following convention when converting
a $p$-form $\omega$ into its components $\omega_{\mu_1\cdots \mu_p}$
by defining
\begin{equation}
\omega=\frac1{p!}\omega_{\mu_1\mu_2\cdots\mu_p}dx^{\mu_1} dx^{\mu_2}
\cdots dx^{\mu_p}.
\end{equation} An example is \begin{equation}
F=\frac1{120} F_{\mu\nu\rho\sigma\tau}dx^\mu dx^\nu dx^\rho dx^\sigma
dx^\tau
\end{equation} for the self-dual five-form $F$.

The zero-th order solution is\begin{align}
ds^2 &= L^2ds^2_{\AdS}+L^2ds^2_{X},&
F&= \frac{2\pi N}{V} \left(\vol_X +\vol_{\AdS}\right). \label{zeroth}
\end{align}
We take the convention $R_{\mu\nu}=-4g_{\mu\nu}$ for the AdS part,
$R_{mn}=4g_{mn}$ for the SE part.
Plugging \eqref{zeroth}
in to the equation of motion of the metric, we get\begin{equation}
4=c\left(\frac {2\pi N}V\right)^2 L^{-8}.
\end{equation}

Let us expand the fluctuation around the zero-th order solution in modes.
One can consistently set to zero  all the modes  which are not
invariant under the $U(1)$ isometries.
We then take the ansatz for  $F_5$ as \begin{equation}
\frac{V}{2\pi N}F_5=\hat e^1\cdots \hat e^5 +
B^a \wedge *k_a - F^I \wedge \hat\omega_I
 +*F^I \wedge \widehat{*\omega_I}+(*B^a)\wedge k_a+f^1\cdots f^5\label{ansatz}
\end{equation} where $k_a=g_{ij} k_a^i dy^j $, $\omega_I$ are three-forms
to be identified shortly, $B^a=B^a_\mu dx^\mu$
and $F^I=F^I_{\mu\nu} dx^\mu dx^\nu/2$.
We will see that this gives
consistent equation of motion in five dimensions.
Note that \begin{equation}
(\omega_I\ \text{elsewhere in the article})=-\frac{2\pi}{V} (\omega_I\ \text{here}).
\end{equation} This definition saves messy factors of powers of $2\pi$.

$F_5$  above  satisfies $F_5=*F_5$ by construction,
because the one-forms  $f^\mu$ and $\hat e^i$ constitutes the zehnbein of the metric.
$dF_5=0$ requires \begin{equation}
d\omega_I = c^a_I\iota_{k_a }\vol_X
\end{equation}
 for some constants $c^a_I$.
We define $k_I\equiv c^a_I k_a$ for brevity. Note also that  \begin{equation}
(k_I\ \text{elsewhere in the article})=2\pi (k_I\ \text{here}).
\end{equation}

Furthermore, we assume $\omega_I$ to be co-closed.
Then,
$dF_5=0$ imposes on $B^a$, $F^I$ the equations
\begin{align}
d(A^a+B^a) &=c^a_I F^I,\label{mix}\\
dF^I&=0,\\
d(*F^I)\wedge *\omega_I &= -(*B^a) \wedge dk_a  + F^I \wedge F^J \wedge
\iota_{k_J} \omega_I    \label{CC}
\end{align}where we kept the fluctuations up to the second order.
Let us define $\omega_a$ by $*dk_a/8$.
One has  $d\omega_a=\iota_{k^a}\vol$
by using the fact\footnote{
One can replace $\partial_i $ by $\nabla_i$ in the definition
of Lie derivative. Thus  $\nabla_i k_j+\nabla_j k_i=0$.
Then  \[
 R_{lj}k^l =R^k_{lkj} k^l=[\nabla_k, \nabla_j] k^k
 =g^{kl}[\nabla_k,\nabla_j]k_l=-g^{kl}\nabla_k\nabla_l k_j-g^{kl}\nabla_j\nabla_k k_l
 =-\nabla^2 k_j.
\]
Hence, for Einstein manifold with $R_{ij}=tg_{ij}$,
we have \[
(*d*dk)_i = \nabla^j(\nabla_i k_j-\nabla_j k_i)=2t k_i.
\] 
}  that we have
$*d*dk= 2t k$ for any Killing vector $k$ in an Einstein spaces
with $R_{ij}=t g_{ij}$.
Then we see, from \eqref{CC},\begin{equation}
d*F^I\wedge \omega_K\wedge *\omega_I
= -8* B^a \wedge\omega_K\wedge*\omega_a
+ F^I \wedge F^J \wedge \omega_K\wedge \iota_{k_J} \omega_I. \label{CC2}
\end{equation}

Another important EOM comes from the Ricci curvature $R_{\mu \hat i} f^\mu \hat e^i$
with one leg in the AdS and one leg in the SE.
While \begin{equation}
R_{\mu \hat i}
 = \frac12 k_{i a} \nabla_\nu (\partial_\mu A^a_\nu-\partial_\nu A^a_\mu)
\end{equation}from \eqref{metricansatzApp},
the right hand side of \eqref{metriceq} is given by\begin{align}
\frac c{24} F_{\mu ....}F_{\hat i}{}^{....}
&=\frac c{24} \left(\frac {2\pi N}V\right)^2L^{-8}\left(
48 B^a_\mu k_{ai}  -
6 (*F^I)_{\mu\nu\rho} (*\omega_I)_{..} F^{J\nu\rho} (\omega_J)_{\hat i}{}^{..}
\right)\\
&=8 B^a_\mu k_{ai} -4(*F^I\wedge F^J)_\mu
\frac{(\omega_I \iota_{e_i} \omega_J) }{\vol_X}.
\end{align}
Thus we get
 \begin{multline}
\frac1{16}  (d*dA^a) \wedge \contract{k_{a}}{k_K}\vol_X=\\
 *B^a\wedge  \contract{k_{a}}{k_K} \vol_X + \frac12
 F^I\wedge F^J\wedge \omega_I \iota_{k_K} \omega_J.  \label{KK}
\end{multline}
where we define $\contract{a}{b}$ for two one-forms $a=a_idx^i$
and $b=b_i dx^i$ by $\contract{a}{b}=a_i b_j g^{ij}$.

From \eqref{CC2} and \eqref{KK} we see $B^a$ are the massive eigenmodes
under Kaluza-Klein expansion,
hence we need to set $B^a=0$ to get the Ansatz
for  the massless fluctuation.
Let us  add the both sides of the equations
\eqref{CC2} and \eqref{KK}, and integrate over the internal manifold $X$.
Using  $\int_X \omega_K\wedge *\omega_a = \int_X k_K k_a\vol_X /8$,
the term including the massive mode $B^a$ vanishes, and
we finally obtain the EOM for
massless fields : \begin{equation}
 d*F^I \int_X (\omega_K\wedge *\omega_I+\frac{1}{16}\contract{k_K}{k_I}\vol_X)
=\frac14 F^I\wedge F^J \int_X \omega_{\{I} \wedge \iota_{k_J} \omega_{K\}}
\end{equation}
where $\{IJK\}=IJK+IKJ+\cdots$  without $1/6$.
The factor which multiplies  $d*dF_I$ exactly reproduces
the combination  $g^{-2}_{IJ}{}^{KK}+g^{-2}_{IJ}{}^{CC}$ which appeared
in ref \cite{currentcorrelators},
where it was derived in a slightly different way.

Let us recapitulate what happens during the detailed calculation.
If we reduce some higher-dimensional form-field theory
on an internal manifold without isometries,
we need to have simultaneously closed and co-closed wavefunctions
in the internal manifold to have a massless field in the non-compact dimensions.
If the metric is the sole dynamical field, then upon reduction
an isometry produces a gauge field through the ansatz \eqref{metricansatz}.
Through the coupling between the metric and five-form field,
the gauge field from $g_{\mu\nu}$ and the gauge field from $F_5$
with co-closed but nonclosed  wavefunctions
get off-diagonal components in the mass matrix,
and precisely one linear combination remains massless for one Killing vector field.
Thus, the number of massless gauge fields
in $\AdS$ is \begin{equation}
d=\ell + b^3,
\end{equation}where $\ell$ is the number of independent Killing vectors
and $b^3$ is the dimension\footnote{
Forms which are closed and co-closed
are automatically invariant under the isometry,
hence the number of harmonic three-forms is the same
as the number of invariant harmonic three-forms.}
of $H^3(X)$.

\section{Triangle anomaly for general toric quivers}\label{generaltoric}
In this appendix we prove the formula
\begin{equation}
c_{IJK}=\frac{N^2 }2 |\det(k_I,k_J,k_L)|
\end{equation} for quiver gauge theories
on the D3-branes probing the tip of a toric Calabi-Yau cone.

Let us denote by $k_I=(1,\vec k_I)$  ($I=1,\ldots,d$)
the toric data of the toric Calabi-Yau manifold.
We set $k_0\equiv k_d$.
One can express the same data using the language of the
$(p,q)$-web, in which  the direction of the $i$-th external leg
is given by $(p_i,q_i)=\vec k_{i}-\vec k_{i-1}$.
The field content
of the corresponding quiver theory is summarized in sec.~\ref{ft},
properties 1, 2 and 3.
Let us consider a linear combination $Q=a^IQ_I$
of the $U(1)$ charges $Q_I$.
Then, the charge of the superpotential under $Q$ is $\sum a_I$
and the charge of the chiral superfields in $\cB_{ij}$ is\begin{equation}
\sum_{K=i}^{j-1} a^K=a_i+a_{i+1}+\cdots+a_{j-1}.
\end{equation} The number $n_{ij}$ of chiral superfields in $\cB_{ij}$ is
given by the intersection number of the two $(p,q)$-legs, that is,
\begin{equation}
n_{ij}=\det (\vec k_{j}-\vec k_{j-1}, \vec k_{i}-\vec k_{i-1}),
\end{equation}while the number $n_V$ of gauge groups
is given by the area of the toric diagram
\begin{equation}
n_V=\sum \det (\vec k_I-\vec k_1,\vec k_{I+1}-\vec k_1).
\end{equation}
Then the triangle anomaly among three $Q$'s is given by
\begin{equation}
\frac1{N^2}c_{IJK}^{\text{CFT}} a^I a^J a^K=n_V (\frac12\sum a^I)^3 +  \sum_{I<J}
n_{IJ} \left( \sum_{K=I}^{J-1} a^K  -\frac12 \sum a^I\right)^3.\label{BZformula}
\end{equation}
This expression follows from the folded quiver picture of
\cite{geo}, and appeared explicitly in the work of Butti and
Zaffaroni \cite{BZ}. In the usual formula we have $1$ instead of
$\sum a^I/2$; we would like to have the triangle anomaly including
the global symmetry usually fixed by $\sum a^I=2$, so we
resurrected that combination.

One can show, by mathematical induction, $c_{IJK}^{\text{CFT}}$
only depends on $k_{I,J,K}$ and not on other $k_{L}$ for $L\ne
I,J,K$.  nor on the number of edges. The proof goes as follows :

Suppose $I,J,K\ne d$ and let us show $c_{IJK}$ is independent of
$k_d$. Consider two toric data, one is the original set $\{
k_1,k_2,\cdots,k_d\}$ and the other is $\{k_1,\cdots,k_{d-1}\}$
without $k_d$. Let us distinguish various quantities for the
latter by adding tilde above, e.g. $\tilde n_V$ and so on. Then we
have two relations
\begin{equation}
n_{I,{d-1}}+n_{I,d}=\tilde n_{I,d-1}
\end{equation}
and \begin{equation}
n_V-n_{d-1,d}=\tilde n_V.
\end{equation} Applying them to the formula \eqref{BZformula},
we obtain\begin{equation}
c_{IJK}^{\text{CFT}} a^I a^J a^K \big|_{a^N=0}
=\tilde c_{IJK}^{\text{CFT}} a^I a^J a^K.
\end{equation} Thus, $c_{IJK}$ for $I,J,K\ne d$ is independent of
$k_d$. Inductively, we can show that $c_{IJK}$ depends only on
$k_I$, $k_J$ and $k_K$.

Hence, we can obtain $c_{IJK}^{\text{CFT}}$
by considering the case of a triangle.
One can easily show that, in this case,\begin{equation}
n_V=n_{IJ}=n_{JK}=n_{KI}=|\det (k_I,k_J,k_K)|.
\end{equation}Plugging in to the formula \eqref{BZformula},
we finally obtain \begin{equation}
c_{IJK}^{\text{CFT}}=\frac {N^2}2 |\det (k_I,k_J,k_K)|.
\end{equation}
It precisely agrees with the result from the supergravity analysis
\eqref{sugraresult}.

\section{More on the charge lattice}\label{more}
We would like to elaborate on the
mathematics of the structure of the charges
of the D3-branes\footnote{The same analysis can be done
for $(d-2)$-branes wrapping $(d-2)$-cycles in a $d$-dimensional
manifold with isometry, since the mixing of the gauge fields
coming from the metric and form-fields is a  generic feature
independent of the self-duality of the form-field, see \cite{currentcorrelators}.
We would like to thank A. Neitzke for raising this question.}.
The case for the toric Sasaki-Einstein manifolds were
analyzed in ref. \cite{LpqQuiver0} mainly from the point of view of the
toric geometry of the cone. We discuss the problem for
arbitrary Einstein manifolds.

Let us denote the space of Killing vectors by $N$,
which can be identified with the Lie algebra of $U(1)^\ell$.
It comes with a natural integral structure by stating that  $k\in N$ is
one of the lattice points if and only if $e^{2\pi k}=id$.
Denote the dual space of $N$ by $M$.
Integral points of $M$ correspond to representations of $U(1)^\ell$.
The Reeb vector $R\in N$ is given
when we endow $X$ with the Sasaki structure.
If $X$ is Sasaki-Einstein, all the toric data $k\in N$ should be on a plane.
The plane
is given by a distinguished element $P\in M$ as $\langle P,k\rangle=1$
where $P$ is the image of $R$ under the identification $M\simeq N$
induced by the metric.

We deliberately used the letters $M$ and $N$ to evoke
the connection with the toric geometry.
Indeed they are precisely $M$ and $N$ lattices of the cone over $X$,
if $\ell =3$.

We only consider the branes which wrap three-cycles invariant
under the action of $U(1)^\ell$. As discussed in section \ref{gg},
two cycles are taken to be equivalent if they form the boundaries
of an invariant four-chain.
Let us call the group of the equivalence classes of such three-cycles as $HG_3(X)$
where $G$ stands for Giant Gravitons.
We also denote the space of linear combinations of $\omega_I$  by $HG^3(X)$,
where $\omega_I$ are closed up to isometry \eqref{closeduptoisometry}.

We have an exact sequence\begin{equation}
0\to H^3(X) \to HG^3(X)\to N\to 0 \label{seq:gaugefields}
\end{equation}
where the second arrow is just the inclusion,
and the third arrow is given by \eqref{closeduptoisometry}.
The exactness of the sequence is also obvious.

Correspondingly, we also have another exact sequence
\begin{equation}
0\to M \stackrel{\iota}{\longrightarrow} HG_3(X)
\stackrel{\pi}{\longrightarrow} H_3(X)\to 0\label{seq:charge}
\end{equation} where we assumed, as before,
that we can take an invariant representative for all $H_3(X)$.
Then, the third arrow $\pi$ is just loosening of the equivalence relation.
The second arrow $\iota$
is a bit tricky to define, so we postpone the
discussion to the end of this section.
In the toric case, the above sequence can be obtained from the
usual sequence \cite{Fulton}\begin{equation}
0\to M\to \mathrm{Div}_T(C(X))\to \mathrm{Pic}(C(X))\to 0.\label{seq:toric}
\end{equation} for the cone $C(X)$ over $X$, where $\mathrm{Div}_T$  denotes
 the group of toric divisors and $\mathrm{Pic}$ is the Picard group.

Two exact sequences have a nice physical interpretation.
First, the relation between various gauge fields are
given by \eqref{seq:gaugefields}. $H^3(X)$ is the wavefunction for
the purely `baryonic' gauge fields, i.e. gauge fields coming from
$F_5$.  The elements of $N$ are the Killing vector fields of $X$,
which give rise to the metric Kaluza-Klein  gauge field.
Formula \eqref{seq:gaugefields} says that the total space of the gauge field is
given by combining the metric and $F_5$ gauge fields, and that
there is generally no gauge fields which come purely from the metric.

Secondly, the sequence \eqref{seq:charge} relates various charges.
Namely, $M$ measures the Kaluza-Klein angular momenta, and
$H_3(X)$ measures the D3-brane charges wrapping various cycles.
The fact that $HG_3(X)$ is the extension of $H_3(X)$ by $M$
tells us that, although we can have excitations with
purely Kaluza-Klein momenta and without D-brane charges, e.g.\ gravitons,
generically any states with D-brane i.e.~`baryonic' charges also have
angular momenta.
It also matches nicely with the result in the
recent works \cite{Kihara:2005nt,Oota:2005mr} which studied
the BPS states with no baryonic charges
and their charge lattice through the analysis of the spectrum
of the Laplacian.
The states without D-brane charges also
appear as the semiclassical strings moving along
the null geodesics. The analysis for $Y^{p,q}$ was
carried out in ref. \cite{Benvenuti:2005cz}.

In the literature on the Sasaki-Einstein/Quiver duality, relatively
little attention is paid to the $M$ part of the charges and the $N$ part
of the gauge fields, so it seems  worthwhile to study further.

Let us now come back to the construction of
the second arrow $\iota$ in \eqref{seq:charge}.
Take an integral basis of Killing vectors $v_a$, $a=1,\ldots,\ell$
of $N$ and take the dual basis $u^a$ in $M$.
The basic idea is first to remove subsets $X_a$ from $X$ so that
$X\setminus X^a$ has a trivial $S^1$ bundle structure
under the action of the vector field $v_a$, second to
take a section of the bundle with its graph $Z_a$,
and finally to set $\iota(v_a)\equiv\partial Z_a$.

The bundle structure is non-trivial, thus
one cannot take a genuine section. The best one can do
is to get a four-chain.
 Then, the boundary of the four-chain is the desired image under $\iota$.
To construct an element $\iota(u^a)$ in $HG_3(X)$ for $u^a$,
first let us denote by $Y^a$  the three-cycle where
the Killing vector $v_a$ degenerates. Define $B^a=(X\setminus Y^a)/U(1)_a$
where $U(1)_a$ is generated by $v_a$.
Then, the orbit of $v_a$ determines a genuine $S^1$ bundle\begin{equation}
S^1\to X\setminus Y^a  \stackrel{p}{\longrightarrow} B^a. \label{bundle}
\end{equation}

Consider the associated vector bundle over $B^a$ obtained by the
fiber $S^1$ by $\C$, and take a generic section of it.
Let the zero locus of the section be given by $t^{ai} \gamma^a_i$
where $\gamma^a_i$ is a two-dimensional submanifold of $b^a$
and $t^{ai}$ is the multiplicity of the zero at $\gamma^a_i$.
Then consider the bundle\begin{equation}
S^1\to X\setminus (Y^a\cup \bigcup_a p^{-1}(\gamma^a_i))  \to
B^a\setminus \bigcup_a\gamma^a_{i}.
\end{equation}  It is a trivial $S^1$ bundle because we removed
$\gamma^a_i$, and we can take a section $Z^a$ of it.

Using $Z_a$, we define the image of $u^a$ by $\iota$
as \begin{equation}
\iota(u^a)\equiv \partial Z^a = Y^a + t^{ai} p^{-1}(\gamma^a_i).
\end{equation}
As before, we assume that we can take $Y^a$ and $\gamma^a_i$
to be invariant under isometries.

The exactness of the sequence \eqref{seq:charge} is now obvious
because the image is the boundary of the four-chain $Z^a$.
Secondly, a D3-brane wrapping on $\partial Z^a$ has
angular momentum $\delta^a_b$ with respect to the isometry $v_b$.
It is because \begin{equation}
\int_{\partial Z^a} \omega_b = \int_{Z^a} d\omega_b
=\int_{Z^a} \iota_{k_b}\vol^\circ =\delta^a_b.
\end{equation}

For the sake of completeness,
we would like to describe the second arrow $\iota$ in \eqref{seq:charge} and
in \eqref{seq:toric} in the toric case.
Let us denote the cone over $X$ by $C(X)$, which is a toric variety.
For $u\in M$, we can take a rational function $\chi^u$ on $C(X)$
satisfying\begin{equation}
v^i\partial_i\chi^u = \sqrt{-1}\vev{u,v}  \chi^u
\end{equation} for $v\in N$, where $\vev{u,v}$
is the natural pairing  between $M$ and $N$. It is unique up to
multiplication by a complex number, since
the torus action is dense in $C(X)$.
Then the image is precisely
the principal divisor $\div(\chi^u)$
determined by $\chi^u$ restricted on $X$,
where the principal divisor $\div(f)$ of a rational function $f$ is \begin{equation}
\div(f)=\sum_\alpha n_\alpha C^\alpha,
\end{equation}  with $C^\alpha$ the loci of the zeros and the poles
of $f$ and with $n_\alpha$ the degree of zeros or the negative of the degree
of poles at $n_\alpha$.

\end{appendix}

\end{document}